\documentclass[aps,pra,reprint,superscriptaddress,nofootinbib]{revtex4-1}

\usepackage[utf8]{inputenc}
\usepackage[T1]{fontenc}
\usepackage{lmodern}
\usepackage{microtype}
\usepackage{mathtools}

\usepackage{placeins}

\usepackage{amsmath}
\usepackage{amsfonts}
\usepackage{graphicx}
\usepackage{hyperref}
\hypersetup{
  colorlinks=true,
  citecolor=blue,
  linkcolor=blue,
  urlcolor=blue
}

\usepackage{mdframed}
\usepackage{comment}
\usepackage[braket,qm]{qcircuit}
\usepackage{multirow}

\newcommand{\cnot}{\textsc{cnot}}

\begin{document}

\title{Resource Efficient Zero Noise Extrapolation with Identity Insertions}

\author{Andre He}
\email{andrehe@lbl.gov}
\affiliation{Physics Division, Lawrence Berkeley National Laboratory, Berkeley, CA 94720, USA}

\author{Benjamin Nachman}
\email{bpnachman@lbl.gov}
\affiliation{Physics Division, Lawrence Berkeley National Laboratory, Berkeley, CA 94720, USA}

\author{Wibe A. de Jong}
\email{wadejong@lbl.gov}
\affiliation{Computational Research Division, Lawrence Berkeley National Laboratory, Berkeley, CA 94720, USA}

\author{Christian W. Bauer}
\email{cwbauer@lbl.gov}
\affiliation{Physics Division, Lawrence Berkeley National Laboratory, Berkeley, CA 94720, USA}

\begin{abstract}
In addition to readout errors, two-qubit gate noise is the main challenge for complex quantum algorithms on noisy intermediate-scale quantum (NISQ) computers.  These errors are a significant challenge for making accurate calculations for quantum chemistry, nuclear physics, high energy physics, and other emerging scientific and industrial applications.  There are two proposals for mitigating two-qubit gate errors: error-correcting codes and zero-noise extrapolation.  This paper focuses on the latter, studying it in detail and proposing modifications to existing approaches.  In particular, we propose a random identity insertion method (RIIM) that can achieve competitive asymptotic accuracy with far fewer gates than the traditional fixed identity insertion method (FIIM).   For example, correcting the leading order depolarizing gate noise requires $n_\cnot+2$ gates for RIIM instead of $3n_\cnot$ gates for FIIM.  This significant resource saving may enable more accurate results for state-of-the-art calculations on near term quantum hardware. 
\end{abstract}

\date{\today}
\maketitle

%%%%%%%%%%%%%%%%%%%%%%%%%%%%%%%%%%%%%%%%%%%%%%%%%%%%%%%
\section{Introduction}

Gate and readout errors currently limit the efficacy of moderately deep circuits on existing noisy intermediate-scale quantum (NISQ) computers~\cite{Preskill2018quantumcomputingin}.  Readout errors can be mitigated with unfolding techniques~\cite{unfolding}.  Two-qubit gates are the most important source of gate noise and the most basic two-qubit gate is the controlled \textsc{not} operation (`\cnot').  One strategy for mitigating these errors is to build in error correcting components into the quantum circuit.  Quantum error correction~\cite{0904.2557,Devitt_2013,RevModPhys.87.307,2013qec..book.....L,Nielsen:2011:QCQ:1972505} is non-trivial because qubits cannot be cloned~\cite{Park1970,Wootters:1982zz,DIEKS1982271}.  As a result, there is a significant overhead in the additional number of qubits and gates requires to make a circuit error-detecting or error-correcting.  This has been demonstrated for simple quantum circuits~\cite{errorcorrecting,PhysRevA.97.052313,Barends2014,Kelly2015,Linke:2017, Takita:2017, Roffe:2018, Vuillot:2018,
  Willsch:2018,Harper:2019}, but is currently infeasible for current qubit counts and moderately deep circuits. 

\begin{figure}[h!]
\begin{mdframed}
\centering
\leavevmode
\large
\Qcircuit @C=0.5em @R=0.8em @!R{
&&&\lstick{\ket{0}}  &\qw  	           &  \ctrl{1}  &  \qw   & \ctrl{1} &  \gate{U_4}  & \meter \\
&&&\lstick{\ket{0}}  & \gate{U_1} & \targ      &  \gate{U_2} &\targ  & \gate{U_3}  &\meter \\
&&&&&&\downarrow&&\\
}\\\vspace{5mm}
\centering
\leavevmode
\Qcircuit @C=0.5em @R=0.8em @!R{
&&&&&\dstick{2n_1+1}&&&&&&&&\dstick{2n_2+1}&&&&\\
 &\qw  	         \gategroup{2}{3}{3}{9}{.5em}{.} \gategroup{2}{11}{3}{17}{.5em}{.}  &  \ctrl{1} &\qw &&\cdots & && \ctrl{1} &  \qw   & \ctrl{1} &\qw & &\cdots & && \ctrl{1} &  \gate{U_4}  & \meter \\
& \gate{U_1} & \targ      &\qw &&\cdots & && \targ &  \gate{U_2} &\targ  &\qw & & \cdots && & \targ & \gate{U_3}  &\meter 
}
\end{mdframed}
\caption{An illustration of identity insertion for a generic controlled unitary operation with two qubits.  The $U_i$ represent unitary matrices and the $n_i$ are non-negative integers.}
\label{fig:circuitillustration}
\end{figure}
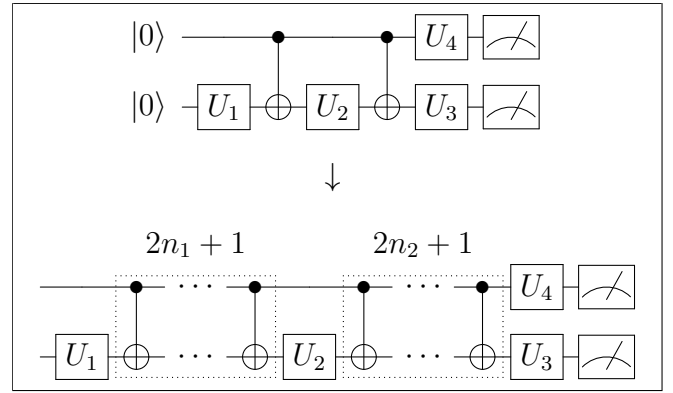

Another strategy for mitigating multigate errors is to find a way to vary the size of the error, measure the result at various values of the error, and then extrapolate to the zero-error result (Zero Noise Extrapolation or ZNE).  With hardware level control of qubit operations, one can enlarge the size of the errors by the gate operation time~\cite{Kandala:2019}.  Such precise hardware level control, however, is often not feasible. Instead, one can try to increase the error algorithmically by modifying the circuit operations.  If the noise model is known, one can insert random Pauli gates to a circuit~\cite{PhysRevX.7.021050}.  For Hamiltonian evolution with some general assumptions on the noise, one can rescale time~\cite{PhysRevLett.119.180509} to amplify the noise by a desired amount.   An approach that does not require knowledge of the noise model is to replace the $i^\text{th}~$\cnot~with 
\begin{align}
\label{eq:insertions}
r_i = 2n_i+1
\end{align}
\cnot~gates, for $n_i\geq 0$.  The focus here is on the \cnot, but the method generalizes to any unitary operation with arbitrary $U^\dag U$ insertions for unitary operation $U$.  Identity insertion is illustrated in Fig.~\ref{fig:circuitillustration}.  Since \cnot$^2$ is the identity, the addition of an even number of \cnot~operations should not change the circuit output, but does amplify the noise.  When $n_i=n$ for all $i$, this is the \textit{fixed identity insertion method} (FIIM).  The application of FIIM was first proposed in Ref.~\cite{Dumitrescu:2018} using a linear fit and an exponential fits were studied in Ref.~\cite{PhysRevX.8.031027}.  Linear superpositions of enlarged noise circuits were also studied in Ref.~\cite{PhysRevLett.119.180509}, which will be similar to our results on higher order fit ZNE with FIIM.   One challenge with FIIM is that it requires a large number of gates.  We propose a new solution to this challenge by promoting the $n_i$ from Eq.~\ref{eq:insertions} to random variables to construct the \textit{random identity insertion method} (RIIM).

This paper is organized as follows.  Section~\ref{sec:linearfiim} reviews linear ZNE in the presence of depolarizing noise.  The RIIM technique is introduced in Sec.~\ref{sec:linearriim}.  The potential of non-linear fits is discussed in Sec.~\ref{sec:nonlinear}.  Sections~\ref{sec:nondepolarizing} and~\ref{sec:shotnoise} extend the discussion to include other sources of quantum noise as well as statistical uncertainties, respectively.  Numerical results with a simple two-qubit circuit and the quantum harmonic oscillator are presented in Sec.~\ref{sec:numerical}.  The paper ends with conclusions and outlook in Sec.~\ref{sec:conclusions}.

%%%%%%%%%%%%%%%%%%%%%%%%%%%%%%%%%%%%%%%%%%%%%%%%%%%%%%%
\section{Linear fit using FIIM in the depolarizing noise model}
\label{sec:linearfiim}

One can build an intuition for the impact of identity insertions analytically using a depolarizing noise model.  In the density matrix formalism, the noisy \cnot~operation between two quibits $k$ and $l$ in the state $\rho$ is given by~\cite{Nielsen:2011:QCQ:1972505}:
\begin{align}
\label{eq:depolarizing}
\rho \rightarrow & \left(1-\sum_{ij}\frac{\epsilon_{ij}^{(kl)}}{16}\right)U^{(kl)}_\text{C} \, \rho \, U^{(kl)}_\text{C}
\nonumber\\
& \qquad +\sum_{i,j = 0}^3\frac{\epsilon_{ij}^{(kl)}}{16} \sigma_i^{(k)}\sigma_j^{(l)}\rho \, \sigma_i^{(k)}\sigma_j^{(l)}
\,,
\end{align}
where $U_\text{\textsc{C}}^{(kl)}$ is the \cnot~operation controlled on qubit $k$ and targeting qubit $l$, $\epsilon_{ij}^{(kl)}\ll 1$ quantifies the amount of noise, and $\sigma_i^{(k,l)}$ is the set of single qubit Pauli gates acting on qubits $k$ and $l$. 

The depolarizing noise model corresponds to the case where all noise parameters $\epsilon^{(kl)} \equiv \epsilon_{ij}^{(kl)}$ are equal to one another, in which case Eq.~(\ref{eq:depolarizing}) becomes
\begin{align}
\label{eq:single_cnot}
\rho\mapsto \left(1-\epsilon^{(kl)}\right) U^{(kl)}_\text{C} \, \rho \, U^{(kl)}_\text{C}+\epsilon^{(kl)} \left(\frac{I_4^{(kl)}}{4}\oplus \rho_{\not kl}\right)
\,, 
\end{align}
where $I_4^{(kl)}$ is the $4\times 4 $ identity matrix on qubits $k$ and $l$, and $\rho_{\not kl}$ is all of $\rho$ aside from the $kl$ qubits.  Equation~(\ref{eq:single_cnot}) has the clear interpretation that with probability $\epsilon^{(kl)}$, $\rho$ is equally likely to be in any of the four possible states: $\ket{kl}\oplus \rho_{\not kl} \in \{\ket{00},\,  \ket{01} ,\, \ket{10},\ket{11}\}\oplus \rho_{\not kl}$.

Suppose that two \cnot~operations are applied sequentially on the same two qubits $k$ and $l$.  The impact on the state is given by:
\begin{align}
\nonumber
\rho&\mapsto \left(1-\epsilon^{(kl)}\right)^2\rho \\\label{eq:repeated2}
&\qquad+\left[1-\left(1-\epsilon^{(kl)}\right)^2\right]\left(\frac{I_4^{(kl)}}{4}\oplus \rho_{\not kl}\right)
\,.\end{align}
Note that in the noiseless limit $\epsilon^{(kl)} \to 0$, Eq.~(\ref{eq:repeated2}) correctly reproduces the fact that the two \cnot~gates form the identity, such that the density matrix is unaffected.
Adding a third \cnot~gate, one finds
\begin{align}
\nonumber
\rho&\mapsto \left(1-\epsilon^{(kl)}\right)^3U^{(kl)}_\text{C} \, \rho \, U^{(kl)}_\text{C} \\\label{eq:repeated3}
&\qquad+\left[1-\left(1-\epsilon^{(kl)}\right)^3\right]\left(\frac{I_4^{(kl)}}{4}\oplus \rho_{\not kl}\right)
\,.\end{align}

Extending the pattern of Eq.~(\ref{eq:single_cnot})-(\ref{eq:repeated3}), applying the same \cnot~ $r_i = 1 + 2 n_i$ times in a row has the same effect as applying it once with the noise amplified by $r_i$
\begin{align}
\nonumber
\rho&\mapsto \left(1-\epsilon_{i}\right)^{r_i}U^{i}_\text{C} \, \rho \, U^{i}_\text{C}  \\\label{eq:repeatedrtimes}
&\qquad+\left[ 1 - \left(1-\epsilon_{i}\right)^{r_i} \right]\left(\frac{I_4^{(kl)}}{4}\oplus \rho_{\not kl}\right)
\,,\end{align}
where the $i^\text{th}$~\cnot~gate connects qubits $k$ and $l$ and to simplify notation, $\epsilon_i=\epsilon^{(kl)}$. The Taylor expansion of Eq.~(\ref{eq:repeatedrtimes}) around $\epsilon_i=0$ to $\mathcal{O}(\epsilon_i)$ is given by
\begin{align}
\label{eq:repeatedexpanded}
\rho&\mapsto (1-r_i \epsilon_i)U^{(kl)}_\text{C} \, \rho \, U^{(kl)}_\text{C}  + r_i \epsilon_i \left(\frac{I_4^{(kl)}}{4}\oplus \rho_{\not kl}\right)
\,.\end{align}
Thus, the action of $r_i$ \cnot~gates in a row is the same as the action of a single \cnot~gate, but with the noise parameter amplified by a factor of $r_i$.  In FIIM, all of the $r_i$ are set to the same value $r$.

Let $M$ be an observable and in a circuit containing $i = 1 \ldots {N_c}$ \cnot~ gates, consider performing a measurement of the expectation value of $M$: $\langle M\rangle=\text{Tr}(M\rho)$.  Using Eq.~(\ref{eq:repeatedrtimes}) results, the expectation value in the presence of depolarizing noise is given by
\begin{align}
\label{eq:obs_r}
\langle M\rangle(r)&=\left(1- r\sum_{i=1}^{N_c} \, \epsilon_i\right)\langle M\rangle_{\rm ex}+ r \sum_{i=1}^{N_c}\, \epsilon_i\langle M\rangle_{{\rm dep}_i}
\nonumber\\
& \qquad +\mathcal{O}\Bigg(\Big(r \sum_{i=1}^{N_c}  \epsilon_i\Big)^2\Bigg)
\,,
\end{align}
where $\langle M\rangle_{\rm ex}$ is the expectation value of the observable in the absence of noise,  $\langle M\rangle_{{\rm dep}_i}$ denotes the expectation value of the observable if the~\cnot~$i$  is replaced with the depolarizing channel, and $r = 1, 3, \ldots$ is the same factor for every~\cnot~gate in the circuit.

From Eq.~\eqref{eq:obs_r}, the noiseless value of the expectation value is given by the measurement at $r = 0$
\begin{align}
\langle M \rangle_{\rm ex} = \langle M \rangle(0)
\,.
\end{align}
Of course, it is not possible to directly perform a measurement at $r = 0$, since all circuits have noise. The idea of ZNE is to extract the noiseless limit by measuring the result of $\langle M\rangle(r)$ for various values of $r$ and extrapolating to the value at $r = 0$.  By construction, a linear fit is effective when the $\mathcal{O}(\epsilon^2)$ terms in Eq.~(\ref{eq:obs_r}) are subdominant (the `linear regime').  In this regime, one expects to remove the dominant $\mathcal{O}(\epsilon)$ terms with a linear fit so that after linear FIIM
\begin{align}
\langle M \rangle_{\rm FIIM} = \langle M \rangle_{\rm ex} +  \mathcal{O}\Bigg(\Big(r_{\rm max} \sum_{i=1}^{N_c}  \epsilon_{i=1}\Big)^2\Bigg)
\,,
\end{align}
where $r_\text{max}$ is the maximum $r$ value so that the circuit is still in the linear regime.

\begin{figure}[h!]
\centering
\leavevmode
\large
\Qcircuit @C=0.5em @R=0.8em @!R{
&&&\lstick{\ket{0}}  &  \ctrl{1}   &\qw  &\targ       & \qw& \meter \\
&&&\lstick{\ket{0}}  & \targ      &  \qw & \ctrl{-1} & \qw   &\meter \\
}
\caption{An illustration of the the simple double gate circuit described in Sec.~\ref{sec:linearfiim}.}
\label{fig:double}
\end{figure}
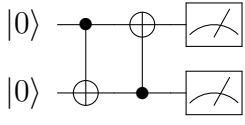

To provide further insight, it is useful to consider an explicit example where the density matrix is easy to compute for arbitrary $r$.  Consider the simple circuit presented in Fig.~\ref{fig:double}.  Due to the small number and simple orientation of gates, this model can be solved completely analytically.   

Letting $r=r_1=r_2 = 1 + 2n$ and $\epsilon=\epsilon_1=\epsilon_2$, applying Eq.~\eqref{eq:repeatedrtimes} to Fig.~\ref{fig:double} results in the following mapping
\begin{align}\nonumber
\rho& \to (1-\epsilon)^{N_cr} U_C^{(12)} U_C^{(21)} \rho \, U_C^{(21)} U_C^{(12)} \\
&\qquad + \left[ 1 - (1-\epsilon)^{N_cr}\right] \frac{I_4}{4}
\,,
\end{align}
where $N_c = 2$ denotes the total number of \cnot~ gates in the circuit and one needs to remember that $r$ is an odd integer
\begin{align}
r = 1 + 2 n 
\,.\end{align}

Thus, starting from the initial state $\ket{00}$ one measures each of the four possible states with probability
\begin{align}
P(\ket{00}) = 1-\frac{3}{4}x \, \qquad P(\ket{ij} \neq \ket{00}) = \frac{x}{4}
\end{align}
where
\begin{align}
\label{eq:xfim}
x_{\rm FIIM}(\epsilon, n) = 1-(1-\epsilon)^{N_c(1 + 2 n)} .
\end{align}

Suppose that one wants to measure $\langle q_0+q_1\rangle$, where $q_i$ is the $i^\text{th}$ qubit in Fig.~\ref{fig:double}. The result of this measurement gives
\begin{align}
\langle q_0+q_1\rangle
&= x_{\rm FIIM}(\epsilon, 0)
\nonumber\\
& = 1 - (1-\epsilon)^{N_c} \nonumber\\
& = N_c \, \epsilon + {\cal O}(\epsilon^2) ,
\end{align}
and is therefore linear in $N_c \,\epsilon$, as expected from Eq.~(\ref{eq:obs_r}). Using \cnot ~noise mitigation, one can remove the linear term in $N_c \, \epsilon$. In the linear FIIM method, one performs the measurement for various values of $n=0,...,n_\text{max}$ and then extrapolates to the value $n = -1/2$ ($r=0$).  A linear fit with these data is a solution to the equation
\begin{align}
\label{eq:leastsquares}
Y=X\beta,
\end{align}
where 
\begin{align}
\label{eq:linearfit}
Y=\begin{pmatrix}x_\text{FIIM}(\epsilon,0)\cr x_\text{FIIM}(\epsilon,1)\cr\vdots\cr x_\text{FIIM}(\epsilon,n_\text{max})\end{pmatrix}\quad X=\begin{pmatrix}0 & 1\cr 1 & 1 \cr \vdots & \vdots\cr n_\text{max} & 1\end{pmatrix}\quad \beta=\begin{pmatrix}\beta_1\cr \beta_0\end{pmatrix}
\,.
\end{align}

The least-squares solution to Eq.~(\ref{eq:linearfit}) is $\beta=(X^TX)^{-1}X^TY$.  This results in the fitted values $\hat{\beta}$:
\begin{align}
\label{eq:beta0}
\hat{\beta}_1&=\frac{\sum_{i=1}^{n}\left(i-\frac{n}{2}\right)x_\text{FIIM}(\epsilon,N_c(1+2i))}{\frac{1}{2}n(n+1)\left(\frac{1}{3}(2n+1)-\frac{n}{2}\right)}\\\label{eq:beta1}
\hat{\beta}_0&=\frac{\sum_{i=1}^{n}\left(\frac{1}{6}n(2n+1)-\frac{ni}{2}\right)x_\text{FIIM}(\epsilon,N_c(1+2i))}{\frac{1}{2}n(n+1)\left(\frac{1}{3}(2n+1)-\frac{n}{2}\right)}
\,.\end{align}
Taylor expanding Eq.~(\ref{eq:beta0}) and~(\ref{eq:beta1}) to $\mathcal{O}(\epsilon^2)$ gives
\begin{align}
\label{eq:beta0e}
\hat{\beta}_1&=2(N_c\epsilon)+(-2n-2+N_c^{-1})(N_c\epsilon)^2+\mathcal{O}(N_{c} \epsilon^3)\\\label{eq:beta1e}
\hat{\beta}_0&=N_c\epsilon+\left(\frac{n(n-1)}{3}-\frac{1-N_c^{-1}}{2}\right)(N_c\epsilon)^2+\mathcal{O}(N_{c} \epsilon^3)
\,.\end{align}
The resulting equation is then 
\begin{align}
\label{eq:FIIM_lin_init}
\langle q_0+q_1\rangle_{\rm FIIM[lin,n_{\rm max}]}=\hat{\beta}_0+\hat{\beta}_1x
\,,\end{align}
where the subscript $\text{FIIM[lin, $n_{\rm max}$]}$ denotes a linear fit performed with the first $n_{\rm max}$ values of $n$.  Inserting Eq.~(\ref{eq:beta0e}) and Eq.~(\ref{eq:beta1e}) into Eq.~(\ref{eq:FIIM_lin_init}) and evaluating at $x=-1/2$  results in
\begin{align}
\label{eq:FIIM_lin}
\langle q_0+q_1\rangle_{\rm FIIM[lin,n_{\rm max}]}&= \left(\frac{2 n_{\rm max}^2 + 4n_{\rm max} + 3}{6}\right) (N_{c} \epsilon)^2 
\nonumber\\
& \qquad \qquad+ {\cal O}(N_{c} \epsilon^3)
\,.
\end{align}

Using more data points makes the extrapolated result worse, rather than better. This can be understood by the fact that using more data points requires more ~\cnot~gates, pushing the measurement into the non-linear regime. One should therefore expect that the error grows with the largest number of~\cnot~gates used, which is given by $r_{\rm max}  N_{c}$. This can clearly be seen by rewriting the result of Eq.~\eqref{eq:FIIM_lin}
\begin{align}
\label{eq:FIIM_lin2}
\langle q_0+q_1\rangle_{\rm FIIM[lin,n_{\rm max}]}\xrightarrow{ r_{\rm \max} \to \infty }& \frac{1}{12} (r_{\rm max} N_{c} \epsilon)^2 
\nonumber\\
& \qquad + {\cal O}(N_{c} \epsilon^3)
\,.
\end{align}
The best result is therefore obtained using a linear fit with 2 points, giving
\begin{align}
\langle q_0+q_1\rangle_{\rm FIIM[lin,1]}&= \frac{3}{2} (N_{c} \epsilon)^2 
+ {\cal O}(N_{c} \epsilon^3)
\,.
\end{align}

A main drawback of linear FIIM is that it requires
\begin{align}
r_{\max} \sum_{i=1}^{N_\cnot}\epsilon_i\sim r_{\max} N_c \, \epsilon \ll 1
\,.
\end{align}
While this works well for circuits for which $N_\cnot \, \epsilon$ is small enough that even after multiplication with $(1+2n)$ it is still a valid expansion parameter, for moderately deep circuits this condition can easily be invalid, in the sense that while one might trust an expansion in $N_\cnot \, \epsilon$, the expansion breaks down for $3 N_\cnot \, \epsilon$ or $5N_\cnot \, \epsilon$. This implies that a linear fit is no longer adequate to extrapolate to the noiseless $(r=0)$ limit.

%%%%%%%%%%%%%%%%%%%%%%%%%%%%%%%%%%%%%%%%%%%%%%%%%%%%%%%
\section{Linear fit using RIIM in the depolarizing noise model}
\label{sec:linearriim}

The main challenge with the linear fit in the FIIM method is that the extrapolated zero noise result is only accurate to ${\cal O}((r_{\rm max} N_{\rm CNOT} \epsilon)^2)$, with $r_{\rm max}$ having to be at least equal to 3. Thus, for deep enough circuits where $3 N_\cnot \, \epsilon \sim 1$ this method completely fails to give an accurate result for the zero-noise extrapolation.  

Since the accuracy of the ZNE depends on the maximum number of \cnot~ gates required, a method that uses less total \cnot~ gates should perform much better.  Instead of inserting the same number of identity operators for every \cnot~gate, suppose instead that identities were randomly inserted.  This gives raise to the \textit{random identity insertion method} (RIIM).  For this approach, one generalizes Eq.~\eqref{eq:obs_r} such that each CNOT gate gets an independent factor $r_i$:
\begin{align}
\nonumber
\langle M\rangle(&r_1,...,r_{N_c})\\\nonumber
&=\left(1- \sum_{i=1}^{N_c} \, r_i \, \epsilon_i\right)\langle M\rangle_{\rm ex}+ \sum_{i=1}^{N_c}\, r_i \, \epsilon_i\langle M\rangle_{{\rm dep}_i}
\\\label{eq:obs_rrim}
& \qquad +\mathcal{O}\Bigg(\Big(\sum_{i=1}^{N_c}  r_i \, \epsilon_i\Big)^2\Bigg)
\,,
\end{align}
Next, the $r_i = 1 + 2 n_i$ in Eq.~\ref{eq:insertions} are promoted to random variables.  For example, one could choose $n_i\sim\text{Poisson}(\nu)$.  As $\nu\rightarrow 0$, a given circuit will have at most one \cnot~gate replaced.  We will show that even in this case, one can still perform a linear fit and thus remove the $\mathcal{O}(\epsilon)$ term with only $N_\cnot+2$ gates instead of $3 N_\cnot$ as in linear FIIM.

Using Eq.~(\ref{eq:repeatedrtimes}) similarly to Eq.~(\ref{eq:obs_r}), one can compute the expectation value of $M$ for RIIM over both the quantum and classical (from sampling $n$) sources of stochasticity:
\begin{align}
\nonumber
\langle \langle M\rangle\rangle(\nu) &=\sum_{n_1=0}^{\infty}\cdots \sum_{n_{\text{N}_\text{CNOTS}}=0}^{\infty} \prod_{i=1}^{\text{N}_c}\Pr(n_i|\nu)\\\nonumber
&\hspace{2mm}\times\Bigg\{ \left[1-\sum_{i}\epsilon_{i}(1+2n_{i})\right]\langle M\rangle_\text{ex}\\\nonumber
&\hspace{10mm}+\sum_{i}\epsilon_{i}(1+2n_{i})\langle M\rangle_{\text{dep}_{i}}\\\label{eq:analytica1}
&\hspace{10mm}+\mathcal{O}\left(\Big(\sum_i (1+2n_i) \epsilon_i\Big)^2\right)\Bigg\}\,
\,.\end{align}
Since each gate is independently sampled, one can replace 
\begin{align}
\sum_{n_i=0}^{\infty} \, \Pr(n_i|\nu) \, n_i = \nu
\,,\end{align}
which immediately reduces Eq.~\eqref{eq:analytica1} to
\begin{align}
\label{eq:analytica1b}
\langle \langle M\rangle\rangle(\rho) &=\left[1-\rho\sum_{i}\epsilon_{i}\right]\langle M\rangle_\text{ex}+\rho\sum_{i}\epsilon_{i}\langle M\rangle_{\text{dep}_{i}}
\nonumber\\
&\hspace{10mm}+\mathcal{O}\left(\Big(\rho\sum_i  \epsilon_i\Big)^2\right)
\,,
\end{align}
where $\rho=1+2\nu$.  Thus, Eq.~\eqref{eq:analytica1b} has the same feature as FIIM, only the integer $n$ is now replaced by the non-integer value $\nu\geq 0$.  By performing measurements at various values of $\nu$ and extrapolating to $\nu = -1/2$, one can extract the noiseless value. However, since the value $\nu$ is not restricted to be integer as in the FIIM case, the expansion does not have to hold for  $3 N_\cnot \, \epsilon$, $5N_\cnot \, \epsilon$, etc., but only for $\rho N_\cnot \, \epsilon$, where one can choose different values of $\nu$ to get a reasonable fit region without making $\rho$ too far from unity.

%%%%%%%%%%%%%%%%%%%%%%%%%%%%%%%%%%%%%%%%%%%%%%%%%%%%%%%
\section{Non-linear fits in the depolarizing noise model}
\label{sec:nonlinear}

So far we have only discussed linear fits and showed that they can eliminate the ${\cal O}(\epsilon)$ noise contribution to a given observable, leaving only quadratic dependence on the noise. In this section we will generalize this result and show that one can in principle eliminate the depolarizing noise to all orders.  This can be done for both the FIIM and RIIM method, which we now discuss in turn.

\subsection{FIIM method}
We begin by revisiting the linear fit in the FIIM method, by writing it in a different way. Starting again from Eq.~\eqref{eq:obs_r}, and setting all $\epsilon \equiv \epsilon_i$ to be equal to one another we can write
\begin{align}
\langle M\rangle(1)&=\langle M\rangle_{\rm ex}+ N_{\cnot} \epsilon\left[ \sum_i \langle M\rangle_{{\rm dep}_i} - \langle M\rangle_{\rm ex} \right]
\nonumber\\
& \qquad +\mathcal{O}(\epsilon^2)
\nonumber\\
\langle M\rangle(3)&=\langle M\rangle_{\rm ex}+ 3 N_{\cnot} \epsilon\left[ \sum_i \langle M\rangle_{{\rm dep}_i} - \langle M\rangle_{\rm ex} \right]
\nonumber\\
& \qquad +\mathcal{O}(\epsilon^2)
\,.\end{align}
One can immediately see that the linear combination 
\begin{align}
\frac{3}{2} \langle M\rangle(1) - \frac{1}{2}\langle M\rangle(3) =  \langle M\rangle_{\rm ex}+\mathcal{O}(\epsilon^2)
\,.\end{align}
This is of course exactly what the linear fit to $r = 0$ using the two points at $r = 1, 3$ would give. 

Generalizing these results one can immediately obtain linear combinations that remove higher order terms in $\epsilon$ as well. This fact has been observed before~\cite{PhysRevLett.119.180509}, and is an application of the Richardson extrapolation~\cite{Richardson, Sidi}. We will still review the results here, since they have not been used in ZNE using CNOT multiplication as a way to increase noise, and will prove useful later. Taking a particular linear combination of the terms with $r = 1, 3, \ldots , r_{\max}$one can eliminate all terms up to ${\cal O}(\epsilon^{n_{\max}+1})$
with
\begin{align}
n_{\max} = \frac{r_{\max} - 1}{2}
\,.
\end{align}
We begin by writing a general linear combination of measurements $\langle M\rangle(r)$ with different values of $r$ and require that this linear combination eliminates all terms up to  ${\cal O}(\epsilon^{n_{\max}+1})$
\begin{align}
\label{eq:fit_equations}
\sum_{n=0}^{n_{\max}}  a(n) \langle M\rangle(1+2n) = \langle M \rangle _{\rm ex} + {\cal O}\left( \epsilon^{n_{\max}+1}\right)
\,.
\end{align}
Ensuring that for any choices of $a(r)$ the coefficient of $\langle M \rangle _{\rm ex}$ is equal to one gives the constraint
\begin{align}
\label{eq:aiSum}
\sum_{n=0}^{n_{\max}} a(n) = 1
\,.
\end{align}
The expression for $\langle M\rangle(r)$ in the depolarizing noise model to all orders in $\epsilon$ can be obtained from Eq.~\eqref{eq:repeatedrtimes} and one finds
\begin{align}
\langle M\rangle(r) = & (1 - \epsilon)^{N_{\rm c} r}  \langle M \rangle _{\rm ex} 
\nonumber\\
& + (1 - \epsilon)^{(N_{\rm c}-1) r} \left[1 - (1 - \epsilon)^{r} \right]  \sum_i \langle M \rangle _{\rm dep_i} 
\nonumber\\
& + (1 - \epsilon)^{(N_{\rm c}-2) r} \left[1 - (1 - \epsilon)^{r} \right]^2 \sum_{i_1,i_2} \langle M \rangle _{\rm dep_{i_1i_2}} 
\nonumber\\
& + \ldots
\nonumber\\
& + \left[1 - (1 - \epsilon)^{r} \right]^{N_{\rm c}} \sum_{i_1,\ldots i_{N_{\rm c}} } \langle M \rangle _{\rm dep_{i_1,\ldots i_{N_{\rm c}} }}
\nonumber\\
=& \langle M \rangle _{\rm ex} - f_{N_{\rm c}}(r, \epsilon) \langle M \rangle _{\rm ex}
\nonumber\\
& +\left[f_{N_{\rm c}}(r, \epsilon) - f_{N_{\rm c}-1}(r, \epsilon)\right] \sum_i \langle M \rangle _{\rm dep_i} 
\nonumber\\
& +\left[f_{N_{\rm c}}(r, \epsilon) - f_{N_{\rm c}-2}(r, \epsilon)\right] \sum_{i_1,i_2} \langle M \rangle _{\rm dep_{i_1i_2}} 
\nonumber\\
& + \ldots
\nonumber\\\label{eq:fullfiim}
& + f_{N_{\rm c}}(r, \epsilon)^{N_{\rm c}} \sum_{i_1,\ldots i_{N_{\rm c}} } \langle M \rangle _{\rm dep_{i_1,\ldots i_{N_{\rm c}} }}
\,,
\end{align}
where 
\begin{align}
f_n(r, \epsilon) = 1 - (1-\epsilon)^{n r}
\,.\end{align}

It is important to remember that the values of $ \langle M \rangle _{\rm ex}$,  $\langle M \rangle _{\rm dep_i} $, $ \langle M \rangle _{\rm dep_{i_1,\ldots i_{N_{\rm c}} }}$ etc. are the results of observables measured in a noiseless circuit, which one does not have access to. This means that when taking linear superposition of the form Eq.~\eqref{eq:fit_equations} the all terms up to ${\mathcal{O}}(\epsilon^n_{\max})$ have to cancel for each line separately. 

This means that the requirement on the coefficients $a(n)$ must satisfy the general equation
\begin{align}
\label{eq:superpositioncondition}
\sum_{n=0}^{n_{\max}}  a(n) f_k(1+2n, \epsilon) =1 + {\cal O}\left( \epsilon^{n_{\max}+1}\right)
\,,
\end{align}
for all values of $k$. 
After some lines of algebra, one can show that this is indeed possible with the coefficients \cite{PhysRevLett.119.180509}
\begin{align}
\label{eq:avals}
a(i) &= \prod_{j=0, j\neq i}^{n_{\max}}\frac{(1+2j)}{2(j-i)} 
\nonumber\\
& = \frac{2^{-2 n_{\max}}}{i!}  \frac{(-1)^i}{1+2i} \frac{(1+2n_{\max})!}{n_{\max}! (n_{\max} - i)!}
\,,
\end{align}
for all $ i \in 1 \ldots n_{\max}$. Note that the coefficient for $i \sim n_{\max} / 2$ is the largest, and satisfies the scaling
\begin{align}
\label{anMax}
{\max}_i[a(i)] \sim a(n_{\max} / 2) \sim \frac{2^{n_{\max}+1}}{n_{\max}}
\end{align}

To summarize, by using values $\langle M \rangle(r)$ with $r = 1, 3, \ldots, r_{\max}$ and taking the linear combination $\sum_{n=0}^{n_{\max}} a(n) \langle M \rangle(1+2n)$, one obtains the noiseless value of the observable up to corrections given by ${\cal O}(\epsilon^{n_{\max} + 1})$.

One alternative approach with a natural interpretation is performing a polynomial fit with degree $n_{\max} - 1$ to measurements of $\langle M \rangle(r)$ with $r = 1, 3, \ldots, r_{\max}$.  A polynomial fit uses the same setup for the linear fit, with Eq.~\eqref{eq:leastsquares}, only now $X$ and $\beta$ are augmented:
\begin{align}
\label{eq:nonlinearfit}
X=\begin{pmatrix}0^{n_\text{fit}} & \hdots & 0 & 1\cr 1^{n_\text{fit}} & \hdots &  1 & 1 \cr \vdots & \vdots\cr n_\text{max}^{n_\text{fit}} & \hdots &n_\text{max} & 1\end{pmatrix}\quad \beta=\begin{pmatrix}\beta_{n_\text{fit}}\cr \vdots \cr \beta_1\cr \beta_0\end{pmatrix},
\end{align}
where $n_\text{fit}$ is the order of the polynomial.  One can show that extrapolating the resulting fit

\begin{align}
\label{eq:FIIM_poly_init}
\langle M\rangle_{\rm FIIM[n_{\rm fit},n_{\rm max}]}=\sum_{i=1}^{n_{\rm fit}}\hat{\beta}_ix^i \, ,
\end{align}
to $x=-\frac{1}{2}$ removes the $\mathcal{O}(\epsilon^{n_{\rm max}})$ component of the depolarizing error when $n_{\rm fit}=n_{\rm max}$.  Both the polynomial fit and the superposition from Eq.~\eqref{eq:avals} give rise to the same linear combinations of the values measured at various values of $r$.  One can show this with some symbolic manipulation:
\begin{align}\nonumber
\langle M&\rangle_{\rm FIIM[n_{\rm fit},n_{\rm max}]}=\sum_{i=0}^{n_\text{fit}}\hat{\beta}_i\left(-\frac{1}{2}\right)^i\\\nonumber
&=\sum_{i=0}^{n_\text{fit}}\sum_{j=0}^{n_\text{max}}((X^TX)^{-1}X^T)_{ij}Y_j\left(-\frac{1}{2}\right)^i\\\nonumber
&=\sum_{j=0}^{n_\text{max}}\left[\sum_{i=0}^{n_\text{fit}}((X^TX)^{-1}X^T)_{ij}\left(-\frac{1}{2}\right)^i\right]Y_j \\\nonumber
&=\sum_{n=0}^{n_\text{max}}\left[\sum_{i=0}^{n_\text{fit}}((X^TX)^{-1}X^T)_{in}\left(-\frac{1}{2}\right)^i\right] \langle M\rangle(1+2n) \\\label{eq:equatecoeffs}
&\equiv\sum_{n=0}^{n_\text{max}}\tilde{a}(n) \langle M\rangle(1+2n) \, .
\end{align}
We have verified that the $\tilde{a}(n)$ in Eq.~(\ref{eq:equatecoeffs}) are equivalent to the $a(n)$ in Eq.~(\ref{eq:avals}).

\subsection{RIIM method}
The RIIM method one uses a different value of $r_i$ for each \cnot~gate.  Applying Eq.~(\ref{eq:repeatedrtimes}) with the full $\epsilon$-dependence leads to the analog of Eq.~(\ref{eq:fullfiim}) from FIIM:
\begin{align}
\langle M\rangle(&r_1, \ldots, r_{N_c}) 
\\
= & \prod_{j}(1 - \epsilon)^{r_j} \Bigg[ \langle M \rangle _{\rm ex} 
\nonumber\\
& + \sum_i  \frac{f_1(r_i, \epsilon)}{(1-\epsilon)^{r_i}}  \langle M \rangle _{\rm dep_i} 
\nonumber\\
& + \sum_{i_1>i_2}  \frac{f_1(r_{i_1}, \epsilon)}{(1-\epsilon)^{r_{i_1}}} \frac{f_1(r_{i_2}, \epsilon)}{(1-\epsilon)^{r_{i_2}}}  \langle M \rangle _{\rm dep_{i_1i_2}} 
\nonumber\\
& +\ldots
\nonumber\\
& +  \sum_{i_1>\ldots > i_{N_{\rm c}}}  \!\!\! \frac{f_1(r_{i_1}, \epsilon)}{(1-\epsilon)^{r_{i_1}}} 
\ldots \frac{f_1(r_{i_{N_{\rm c}}}, \epsilon)}{(1-\epsilon)^{r_{i_{N_{\rm c}}}}}  \langle M \rangle _{\rm dep_{i_1\ldots i_{N_{\rm c}}}} \nonumber
\,.\end{align}
To eliminate all terms up to order $\epsilon^{n_{\max}}$, one needs to include all possible combinations of $r_1,  \ldots, r_{N_c}$ with $\sum_i r_i = N_{\rm c} + 2 n_{\max}$. To write a generic solution we require a bit of new notation. Denote by $O(\{ e_1, \dots e_n\})$ the sum of all operators with the $r_i$ given by permutations of $1$ and the various values of $e_i = 3, 5, 7, \ldots$. So 
\begin{align}
O(\{\}) &= O(1, \ldots, 1)
\\
O(\{e_1\}) &= O(e_1, 1, \ldots, 1) + O(1, e_1,  \ldots, 1) + \ldots
\nonumber\\
O(\{e_1, e_2\}) &= O(e_1, e_2, 1, \ldots, 1) + O(e_1, 1, e_2,  \ldots, 1) + \ldots
\nonumber
\end{align}
and so on. 

To eliminate all terms up to $\epsilon^{n_{\max}}$ one include all operators $O(\{ e_1, \dots e_n\})$ with $\sum_i e_i \leq 2 n_{\max} + n$, each with its own coefficient. One  then determine the coefficients by demanding that all terms up to $\epsilon^{n_{\max}}$ vanish. So for example, to eliminate the linear term in $\epsilon$ one include 
include the operator $O(\{\}) $ and $O(\{3\})$. 
Solving the equations
\begin{align}
a_{\{\}} O(\{\}) + a_{\{3\}} O(\{3\}) = 0 + {\cal O}(\epsilon^2)
\end{align}
with 
\begin{align}
a_{\{\}} = 1 - a_{\{3\}} N_{\rm c}
\end{align}
Solving this equation, one finds
\begin{align}
a_{\{3\}} = - \frac{1}{2}
\,,\end{align}
which again reproduces the result of the linear fit discussed in Section~\ref{sec:linearriim}. 
To eliminate the linear and quadratic term in $\epsilon$ once includes the operators $O(\{\})$, $O(\{3\})$, $O(\{5\})$ and $O(\{3,3\})$, and solves the equation
\begin{align}
& a_{\{\}} O(\{\}) + a_{\{3\}} O(\{3\}) + a_{\{5\}} O(\{5\}) + a_{\{3,3\}} O(\{3,3\}) 
\nonumber\\
& \qquad \qquad = 0 + {\cal O}(\epsilon^3)
\,,
\end{align}
again with the constraint
\begin{align}
a_{\{\}} = 1 - a_{\{3\}} N_{\rm c} - a_{\{5\}} N_{\rm c}  - a_{\{3, 3\}} {N_c \choose 2}
\,.
\end{align}
Solving the resulting set of equations gives
\begin{align}
a_{\{3\}} = - \frac{N_{\rm c} + 4}{4} \,, \qquad a_{\{5\}} = \frac{3}{8} \,, \qquad a_{\{3, 3\}} = \frac{1}{4}
\,.\end{align}

While we have not been able to derive a closed form expressions for the coefficients yet, we report valid choices for the various coefficients with $n_{\max} = 1, 2, 3, 4$ in Table~\ref{RIIMCoefficients}. These results allow to remove depolarizing noise with corrections arising at $\epsilon^{n_{\max} + 1}$ using $N_{\rm c} + 2 n_{\max}$ gates. This should be compared with the FIIM method where the same noise reduction requires $(2 n_{\max} + 1) N_{\rm c}$ gates.

For relatively shallow circuits, one could feasibly perform the measurements for all permutations required for $O(\{e_1, ..,e_n\})$.  For example, to remove the $\mathcal{O}(\epsilon)$ error, one would need to perform $N_c+1$ sets of measurements.  However, this quickly becomes impractical.  This can be circumvented by randomizing: for each measurement that goes into $O(\{e_1, ..,e_n\})$, randomly pick one of the $N_{\{e_1,...,e_n\}}$ operations.

Table~\ref{table:fiimvriim} provides an overview of the gate count requires for FIIM and RIIM in the removal of depolarization noise at a given order in $\epsilon$.

\begin{table}[h!]
\begin{center}
\begin{tabular}{c|c|c}
Method & Remainder & \# of \textsc{CNOT}s \\
\hline
\multirow{1}{*}{FIIM} &  $\mathcal{O}(\epsilon^n)$ & $(2n-1)n_\cnot$\\
\multirow{1}{*}{RIIM} &  $\mathcal{O}(\epsilon^n)$ & $n_\cnot+2(n-1)$\\
\hline
\end{tabular}
\end{center}
\caption{A comparison of the gate count needed for a given order of depolarization error correction for FIIM and RIIM.}
\label{table:fiimvriim}
\end{table}

%%%%%%%%%%%%%%%%%%%%%%%%%%%%%%%%%%%%%%%%%%%%%%%%%%%%%%%
\section{Beyond the depolarizing noise model}
\label{sec:nondepolarizing}

Equation~(\ref{eq:depolarizing}) introduced the full Krauss representation of a noisy~\cnot~gate.  Let $\epsilon_{ij}=\epsilon+\delta_{ij}$.  The depolarizing error model is the case where $\delta_{ij}=0$ and is what has been considered thus far.  In reality, there will be some non-zero $\delta_{ij}$, though the non-depolarizing error has been less studied in the literature and less-characterized on current hardware platforms.  While the methods studied in the previous sections are able to suppress the depolarizing error to $\mathcal{O}(\epsilon^{n_{\max}})$, they do not remove the $\mathcal{O}(\delta)$ term. This means that it is not useful to go beyond $\mathcal{O}(\epsilon^2)$, unless $\delta < \epsilon^2$.

There are many other sources of noise, important examples being amplitude damping and decoherence noise. The latter can be well-approximated as an exponential random variable per operation, where the gate has some fidelity (time constant) and requires some finite time to perform.  We leave the study of such noise to future investigations, but we anticipate that methods similar to those studied here can be used to remove noise other than depolarizing noise as well. In fact, in~\cite{PhysRevLett.119.180509} it was argued that similar methods also apply to amplitude damping noise. 
%%%%%%%%%%%%%%%%%%%%%%%%%%%%%%%%%%%%%%%%%%%%%%%%%%%%%%%
\section{Statistical Uncertainty}
\label{sec:shotnoise}

All results presented so far were in the limit where one can measure the value of an observable with arbitrary precision. This is of course not true, since any measurement on a quantum computer is probabilistic in nature, such that most measurements have a statistical uncertainty associated with them, which depends inversely on the square root of the number of runs used to perform the measurement.

Using the results of the previous sections, one can quantify the impact of the statistical uncertainty. Recall that the noiseless value $\langle M \rangle _{\rm ex}$ is obtained by taking linear combinations of measurements with different values of $r$, and that in the limit of zero statistical uncertainty the final uncertainty on the noiseless value is given by the maximum of $\delta$ and $\epsilon^{n_{\max}+1}$. In the presence of statistical uncertainty, each measurement of $\langle M \rangle _{\rm ex}(r)$ can only be determined up to a statistical uncertainty
\begin{align}
\Delta(r) \sim \frac{1}{\sqrt{n_{\rm meas}}}
\,,\end{align}
where $n_{\rm meas}$ denotes the number of measurements that are performed in the measurement of each value $\langle M \rangle(R)$. 
Adding the various contributions arising from the linear superposition in quadrature, one finds that the error from statistical uncertainties is given by
\begin{align}
\Delta_{\rm stat} & = \frac{1}{\sqrt{n_{\rm meas}}} \sqrt{\sum_{n=0}^{n_{\max}} [a(n)]^2}
\nonumber\\
& \sim \frac{1}{\sqrt{n_{\rm meas}}} \frac{2^{n_{\max}}}{n_{\max}}
\,,
\label{eq:deltastat}
\end{align}
where the last line is only true in the limit of large $n_{\rm max}$, since we have used that the sum is dominated by its largest values, given in Eq.~\eqref{anMax}.

This means that the final uncertainty in the FIIM and RIIM methods are given by
\begin{align}
\Delta_{\rm FIIM/RIIM}[\epsilon, \delta; n_{\rm max}, n_{\rm meas}] \sim \max \left[ \delta, \epsilon^{n_{\max}}, \Delta_{\rm stat} \right]
\label{eq:deltaFIIMRIIM}
\end{align}

%%%%%%%%%%%%%%%%%%%%%%%%%%%%%%%%%%%%%%%%%%%%%%%%%%%%%%%
\section{Numerical results}
\label{sec:numerical}

We use \texttt{qiskit}~\cite{qiskit} to simulate the quantum circuits described below and demonstrate FIIM and RIIM.  Section~\ref{sec:simplecircuit} studies the simple \cnot~only circuit from Fig.~\ref{fig:double} and Sec.~\ref{sec:hamiltonian} examines a more complicated case of time evolution for the quantum simple harmonic oscillator. 

\subsection{Simple Circuit}
\label{sec:simplecircuit}

The simple circuit shown in Fig.~\ref{fig:double} was particularly useful because of its analytical tractability.  In particular, because one can compute the expectation values analytically, it is possible to consider the $n_\text{shots}\rightarrow\infty$ limit.  In this section, we use a slight modification of this simple circuit, which uses 4 \cnot~ gates, which are started in the initial state $\ket{10}$. 
\begin{figure}[h!]\centering
\leavevmode
\large
\Qcircuit @C=0.5em @R=0.8em @!R{
&&&\lstick{\ket{1}}  &  \ctrl{1}   &\qw  &\targ  &\qw   &  \ctrl{1}   &\qw  &\targ       & \qw& \meter \\
&&&\lstick{\ket{0}}  & \targ      &  \qw & \ctrl{-1}  &\qw  & \targ      &  \qw & \ctrl{-1}  & \qw   &\meter \\
}
\caption{A simple circuit with 4 \cnot~ gates used in this section.}
\label{fig:4cnot}
\end{figure}
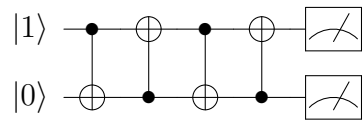
In the noiseless limit, the final state is given by $\ket{11}$. Four gates are used in order to demonstrate the potential for removing depolarization errors up to $\epsilon^5$, and we use a different initial state such that decoherence, discussed later in the section, is not driving the result towards the final expectation.

\begin{figure}[h!]
\centering
\includegraphics[width=0.50\textwidth]{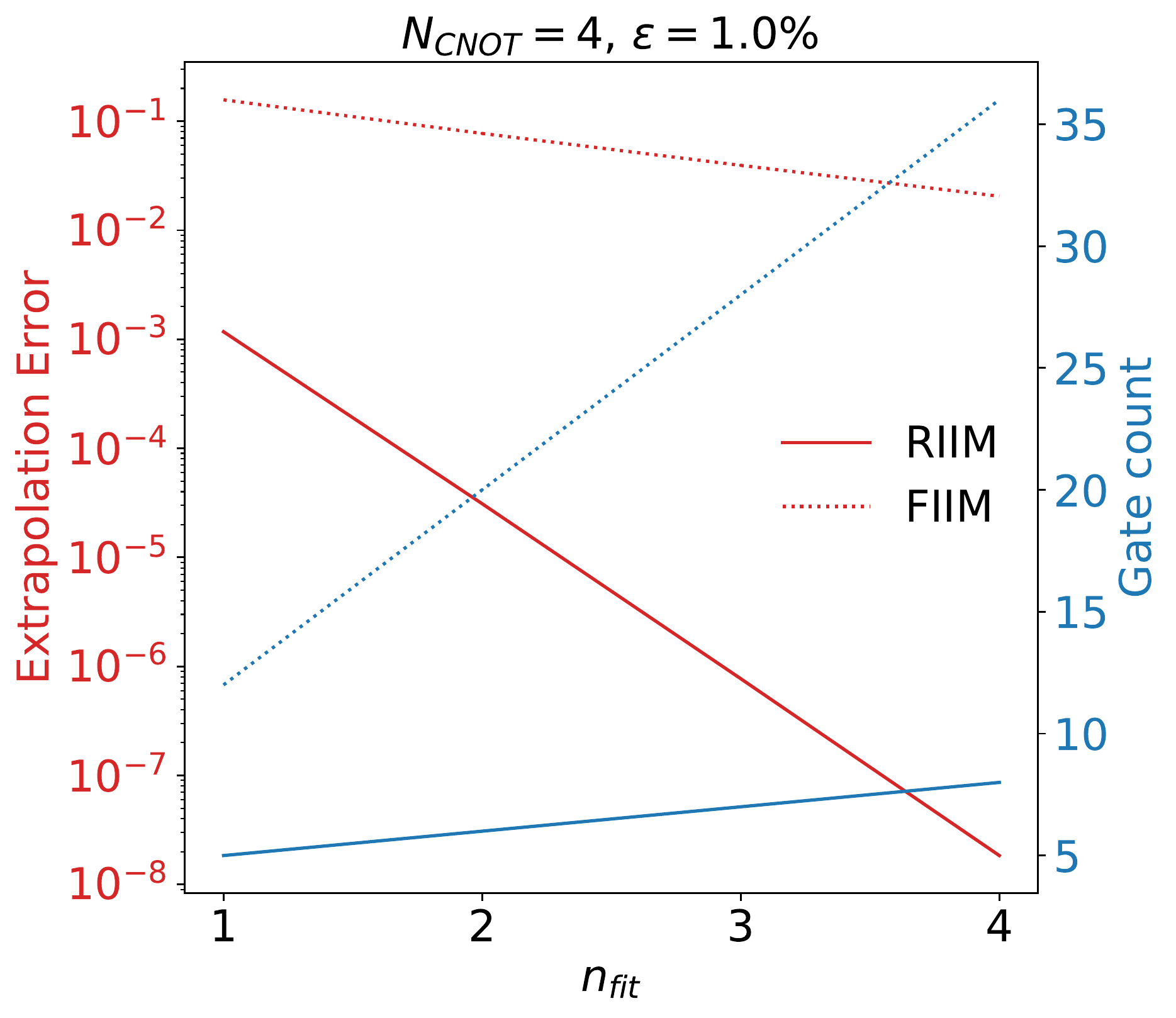}
\caption{Numerical results based on the higher-order fits described in Sec.~\ref{sec:nonlinear} using the four \cnot~gate version of the model presented in Fig.~\ref{fig:double}.  The horizontal axis is the order of the depolarizing error that is being removed.  The left axis is the error on $\langle \sum_{i=0}^{N_c} q_i\rangle$ as $\epsilon$ is extrapolated to zero.  The right axis is the number of gates requires to make the correction.  Only depolarizing noise is considered and $n_\text{shots}=\infty$.}
\label{fig:shotnoise}
\end{figure}

 Fig.~\ref{fig:shotnoise} illustrates the scaling of the error and gate count for RIIM and FIIM for this circuit.  
As desired, the error decreases with the order of the error correction.  The number of qubits required for RIIM is much lower than FIIM for a fixed order of error correction.  For example, correcting the $\mathcal{O}(\epsilon^4)$ requires $8$ total gates for RIIM but FIIM requires $36$.  In fact, for a fixed correction order, the coefficient of the subleading depolarizing error is also smaller for RIIM than for FIIM.

\texttt{qiskit} can be used to study the impact of other sources of noise, such as thermal relaxation.   A full noise model from the IBMQ device is used, which includes depolarizing and decoherence errors.
\begin{figure}[h!]
\centering
\includegraphics[width=0.5\textwidth]{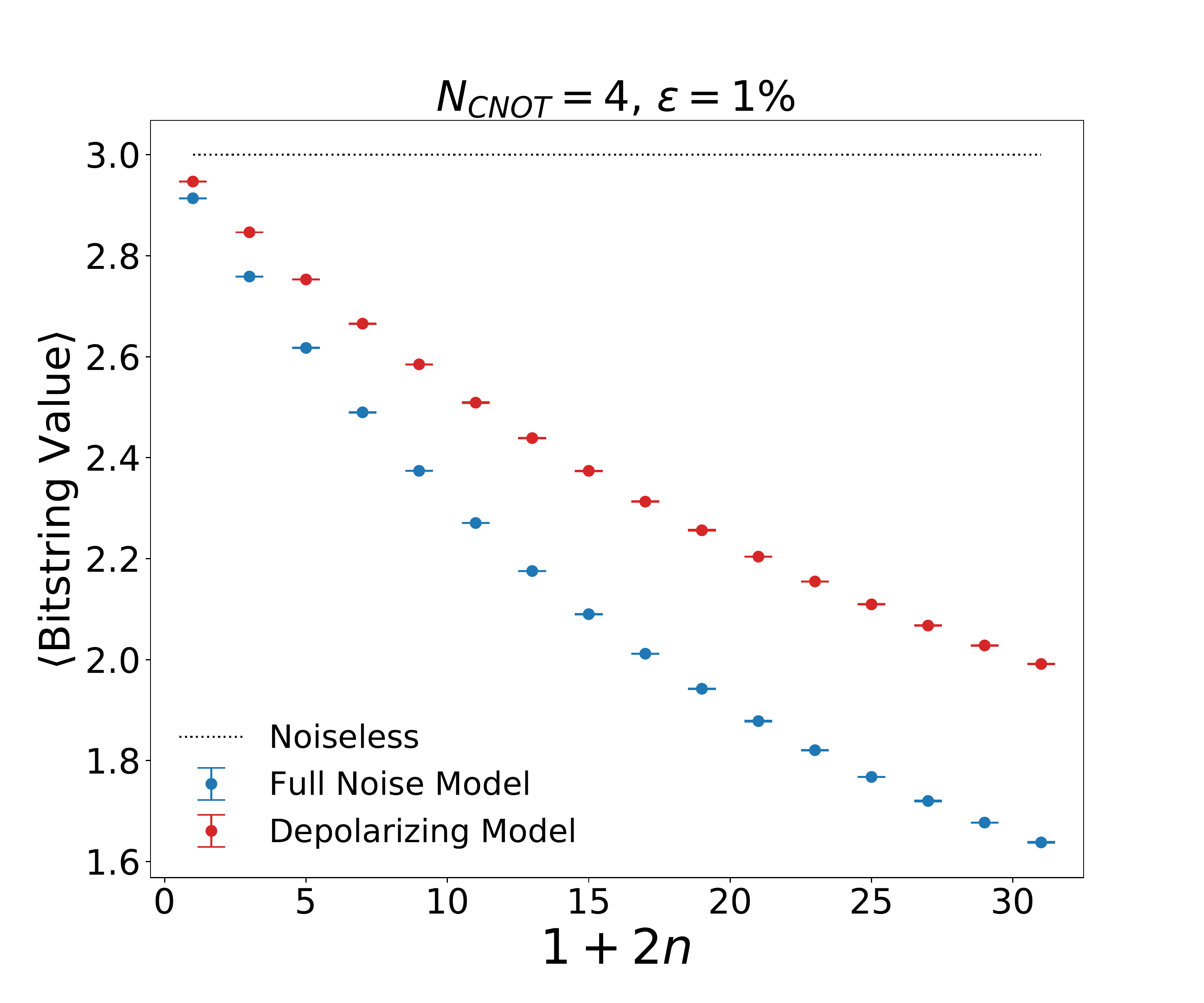}
\caption{Numerical results from simulating the 4-\cnot~circuit in noiseless and noisy simulators using \texttt{qiskit}. The vertical axis shows the expectation value of the measured observable. The horizontal axis displays $r$, or $1+2n$. Noisy simulations include both full and purely depolarizing cases. The number of shots for each point is $10^7$, with a standard deviation of $10^{-3}$. }\label{fig:data_2cx}
\end{figure}

In Fig.~\ref{fig:data_2cx}, we show the result where the measured observable is the expected value of the output string, converting from binary numbers to integers ($00\rightarrow 0, 01\rightarrow 1, 10\rightarrow 2, 11\rightarrow 3$).  In the noiseless limit, the expectation value is 3, corresponding to $\ket{11}$
Fixed identity insertions (but no corrections yet) are applied up to $r_\text{max}=31$. The observable decays at a quicker rate in the case with the full noise model as expected, as the circuit feels the effect of thermal relaxation (which drives the system towards the $\ket{00}$ state) as well as the depolarizing noise, which drives the system to the completely mixed state. 

\begin{figure}[h!]
\centering
\includegraphics[width=0.5\textwidth]{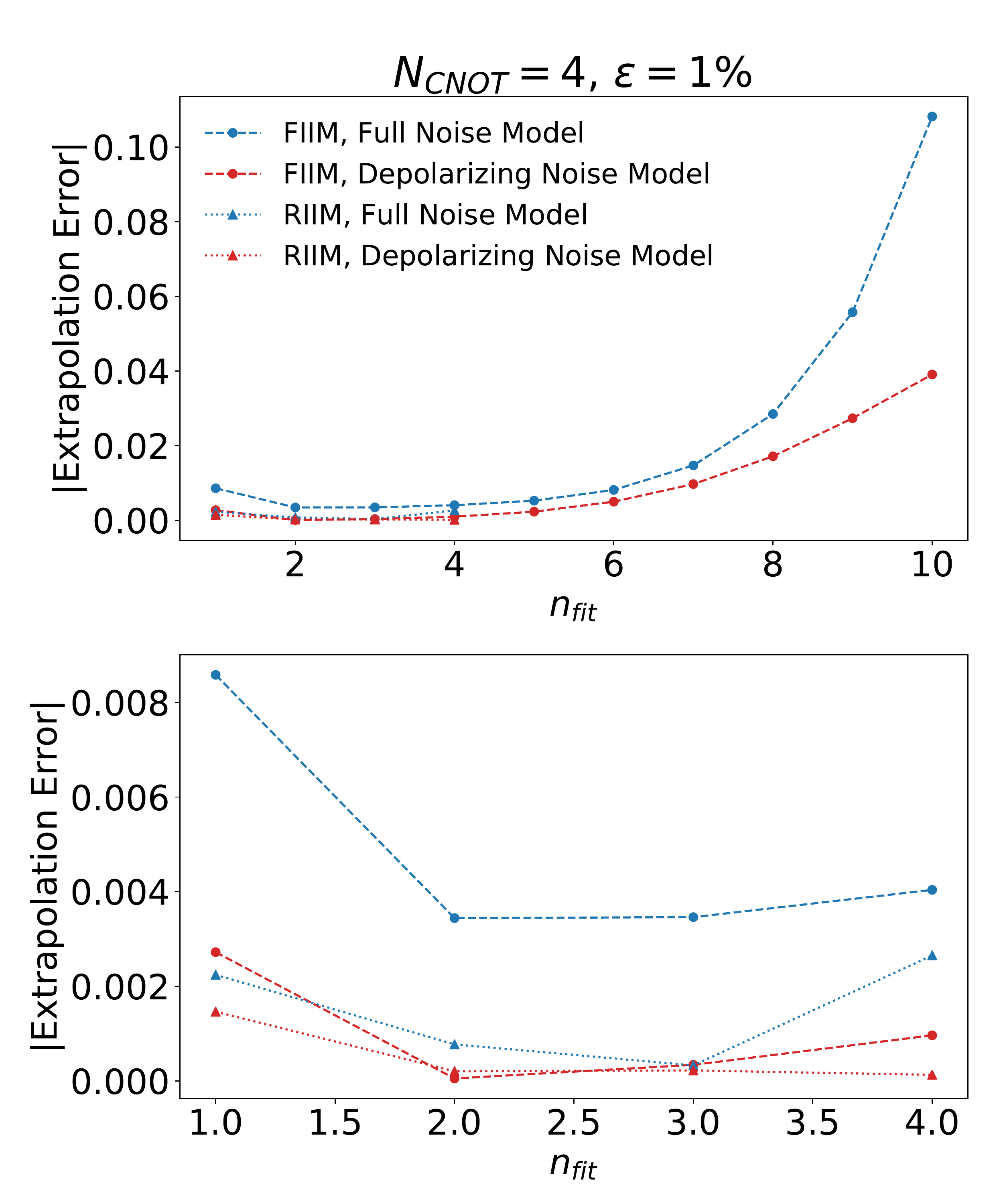}
\caption{Numerical results from simulating the 4-$\cnot$ circuit and applying FIIM and RIIM extrapolation in a noisy simulator. The lower plot displays a reduced range of the data in the upper plot for detail. Extrapolation error is 0 for noiseless simulation. The number of shots for each point is $10^7$.}
\label{fig:err_2cx}
\end{figure}

Fig. \ref{fig:err_2cx} compares the extrapolation error obtained from FIIM and RIIM under the action of full and purely depolarizing noise models. The extrapolation error for both FIIM and RIIM are higher in the case of a full noise model which has non-depolarizing elements. RIIM performs as well or better than FIIM in both noise models. The minimum extrapolation error is achieved at $n_\text{fit}=2$. This can be understood in the context of Eq. \eqref{eq:deltaFIIMRIIM} in Section \ref{sec:shotnoise} with the parameters of $\epsilon=1\%$ and $n_{\rm meas} = 10^7$. At $n_\text{fit}=1$, the dominant error $\Delta_\text{FIIM/RIIM}$ is determined by $\epsilon$ rather than the statistical uncertainty. However, as $n_\text{fit}=2$, the statistical error $\Delta_{\rm stat}$ begins to exceed $\epsilon^{2}$, and by $n_\text{fit} = 3$ the dominant error becomes the statistical error, which prevents further reduction of extrapolation error and leads the the exponential scaling of the error as $n_\text{fit}$ is increased further.  Note that RIIM is only used to eliminate errors up to $\mathcal{O}(\epsilon^4)$, as the circuit only contains 4 $\cnot$s. 

\subsection{Hamiltonian Evolution}
\label{sec:hamiltonian}

Trotterized time evolution is a useful technique for the simulation of Hamiltonians on digital quantum computers. For the one-dimensional simple harmonic oscillator Hamiltonian, time evolution is given by
\begin{align}
\ket{\psi(t)}= e^{-iHt}\ket{\psi(0)}
\,,
\end{align}
where 
\begin{align}
\label{eqOld:HOhamiltonian}
H  = \frac{1}{2}(\hat{x}^2 + \hat{p}^2) \equiv H_x + H_p
\,.
\end{align}

\begin{figure}[h!]
\begin{mdframed}
\centering
\leavevmode
\large
\Qcircuit @C=0.3em @R=0.1em @!R{
&\targ & \gate{Rz} & \targ &\multigate{1}{U_{QFT}} & \qw &\targ & \gate{Rz} &  \targ & \multigate{1}{U_{QFT}^\dag} & \qw\\
&\ctrl{-1} & \qw & \ctrl{-1} &\ghost{U_{QFT}}  & \qw&\ctrl{-1} & \qw & \ctrl{-1} & \ghost{U_{QFT}^\dag} & \qw\\
}
\centering
\leavevmode
\end{mdframed}
\caption{Circuit diagram for a single Trotter step of the time evolution of the harmonic oscillator hamiltonian for two qubits. The total number of \cnot ~operations for the quantum Fourier transform $U_{\rm QFT}$ on 2 qubits is 5, giving a total of 14 \cnot operations. However, one \cnot ~operation from each of the $U_{\rm QFT}$ is cancelling a \cnot~from the rest of the circuit, giving a total of 10 \cnot~operators per Trotter step.}
\label{fig:circuitillustration2}
\end{figure}
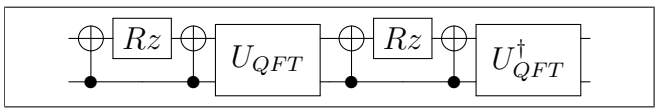

The Hamiltonian in Eq.~\eqref{eqOld:HOhamiltonian} can be implemented on a digital quantum computer by discretizing the possible values of $x$ to be $-x_\text{max},-x_\text{max}+\delta_x,\cdots,x_\text{max}-\delta_x,x_\text{max}$, where $\delta_x=2x_\text{max}/(2^{n_\text{qubits}}-1)$ and $n_\text{qubits}$ is the number of qubits.  This system has been recently studied in the context quantum field theory as a benchmark $0+1$ dimensional non-interacting scalar field theory~\cite{Jordan:2017lea,Jordan:2011ci,Jordan:2011ne,Jordan:2014tma,10.5555/3179430.3179434,PhysRevLett.121.110504,Macridin:2018oli,Klco:2018zqz}.  As discussed in these studies, the momentum operator $\hat{p}^2$ can be effectively implemented with quantum Fourier transforms.  Since $[H_x, H_p] \neq 0$, one can approximate the time evolution of the Hamiltonian by using the first-order Suzuki-Trotter expansion~\cite{10.2307/2033649,Suzuki1976,1976PThPh..56.1454S}:
\begin{align}
\label{eqOld:trotterapprox}
e^{-i(H_x + H_p)t} &\approx \left[e^{-iH_x\frac{t}{n}}e^{-iH_p\frac{t}{n}}\right]^n
\nonumber\\
& \equiv \left[U^{(H)}_{n}(t/n)\right]^n
\end{align}
The approximation in Eq.~\eqref{eqOld:trotterapprox} can be efficiently represented as a quantum circuit block which is repeated $n$ times to the desired number of Trotter steps, as illustrated in Fig.~\ref{fig:circuitillustration2}.

Time evolution of the ground state of the Harmonic oscillator gives
\begin{align}
\ket{\psi_0(t)} = e^{-iHt}\ket{\psi_0(0)} = e^{-i E_0 t} \ket{\psi_0(0)} ,
\end{align}
where $E_0 = 1/2$. Thus, the time evolution produces a pure phase and one finds 
\begin{align}
\ip{\psi_0(0)}{\psi_0(t)} = 1
\,.
\end{align}
The ground state of the harmonic oscillator is a Gaussian distribution in the variable $x$, which can be generated through the action of a unitary circuit on the state $\ket{0}$. $U_{\rm State}$ is implemented with 2 $\cnot$ gates. 
\begin{align}
\ket{\psi_0(0)} = U_{\rm State} \ket{0}
\,.
\end{align}
Thus, the overlap can be written as
 \begin{align}
\label{eqOld:diagnostic}
\lim_{n \to \infty} \bra{0}U_\text{State}^\dag\left[U^{(H)}_{n}(t/n)\right]^n U_\text{State}\ket{0} = 1
\,.
 \end{align}
For finite values of $n$ the deviation of the overlap from unity will grow with time $t / n$ and one achieves higher accuracy for larger $n$
\begin{align}
 \bra{0}U_\text{State}^\dag\left[U^{(H)}_{n}(t/n)\right]^n U_\text{State}\ket{0} = 1 + {\cal O}\left(t^2/n^2\right)
 \,.\end{align}
On the other hand, more Trotter steps requires deeper circuits, and therefore larger errors from the gate noise, in particular the \cnot~noise.

We choose to simulate the harmonic oscillator with a total of 2 qubits, corresponding to 4 discrete values of $x$. In this case the  \cnot~count is given by
\begin{align}
N_{\rm c} = 4 + 10 \, n
\,.
\end{align}
The accuracy of the approximation increases with the number of Trotter steps $n$.  FIIM has been used to increase the accuracy of Trotterized simulation of the time evolution of Hamiltonians, but is less accurate when the depth of a single Trotter step becomes too large, as introducing three or more times as many \cnot~operations as there are in the nominal circuit does not allow for the accurate extrapolation of the observable~\cite{klco2019su2}.

\begin{figure}[h!]
\centering
\includegraphics[width=0.52\textwidth]{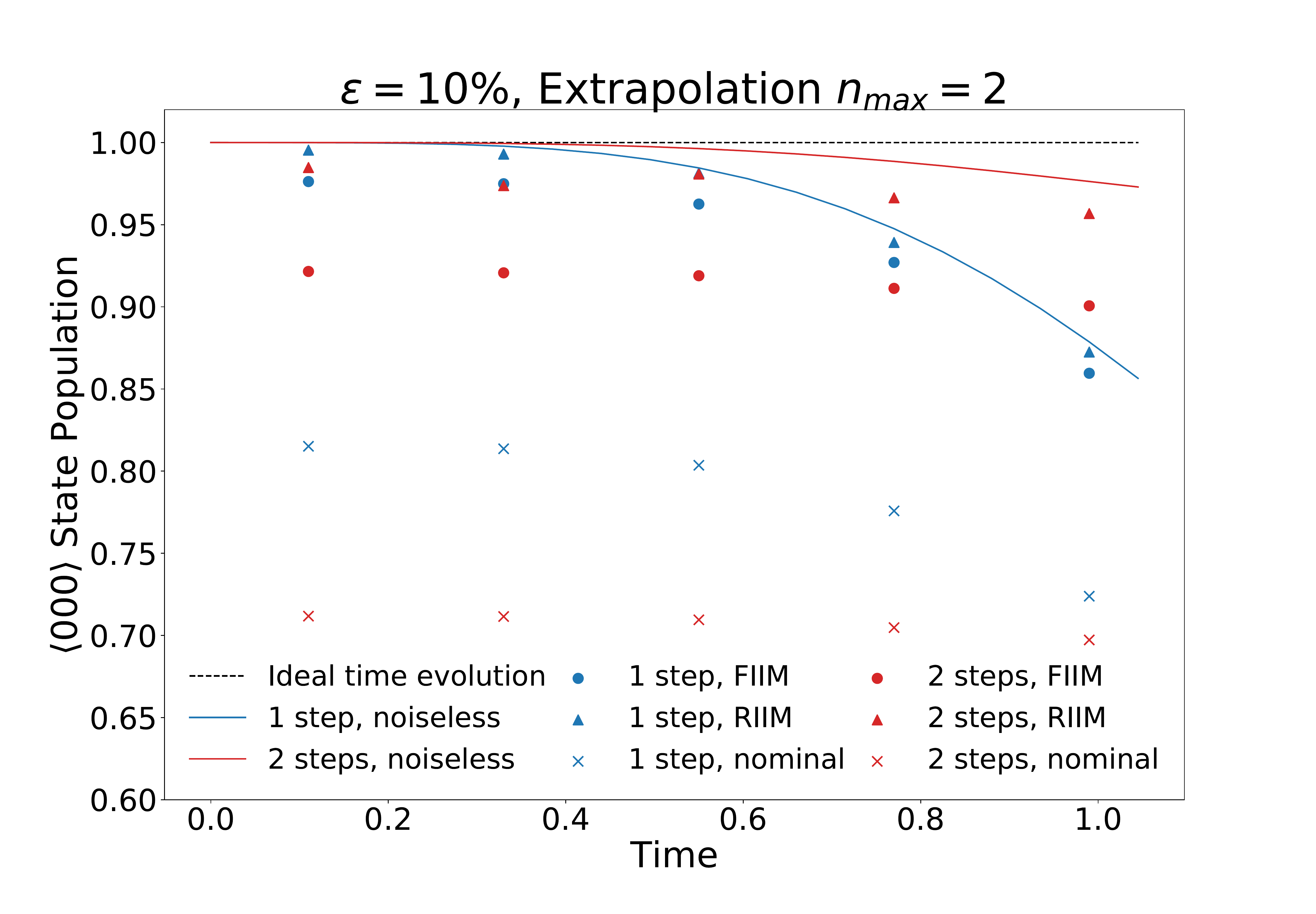}
\caption{The expectation value of the observable in Eq.~\eqref{eqOld:diagnostic} as a function of time for various numbers of Trotter steps and noise mitigation techniques. The purely depolarizing noise model is used for this simulation. In the absence of Trotterization error, the observable should be unity, independent of $t$. }
\label{fig:trotter_1}
\end{figure}

Fig.~\ref{fig:trotter_1} presents the result of one and two Trotter steps, corrected with RIIM and with FIIM up to $\mathcal{O}(\epsilon^2)$. For both one and two steps, the RIIM extrapolations are closer to the noiseless lines than the FIIM extrapolations, indicating that the RIIM error is smaller than the FIIM one.  

\begin{figure}[h!]
\centering
\includegraphics[width=0.50\textwidth]{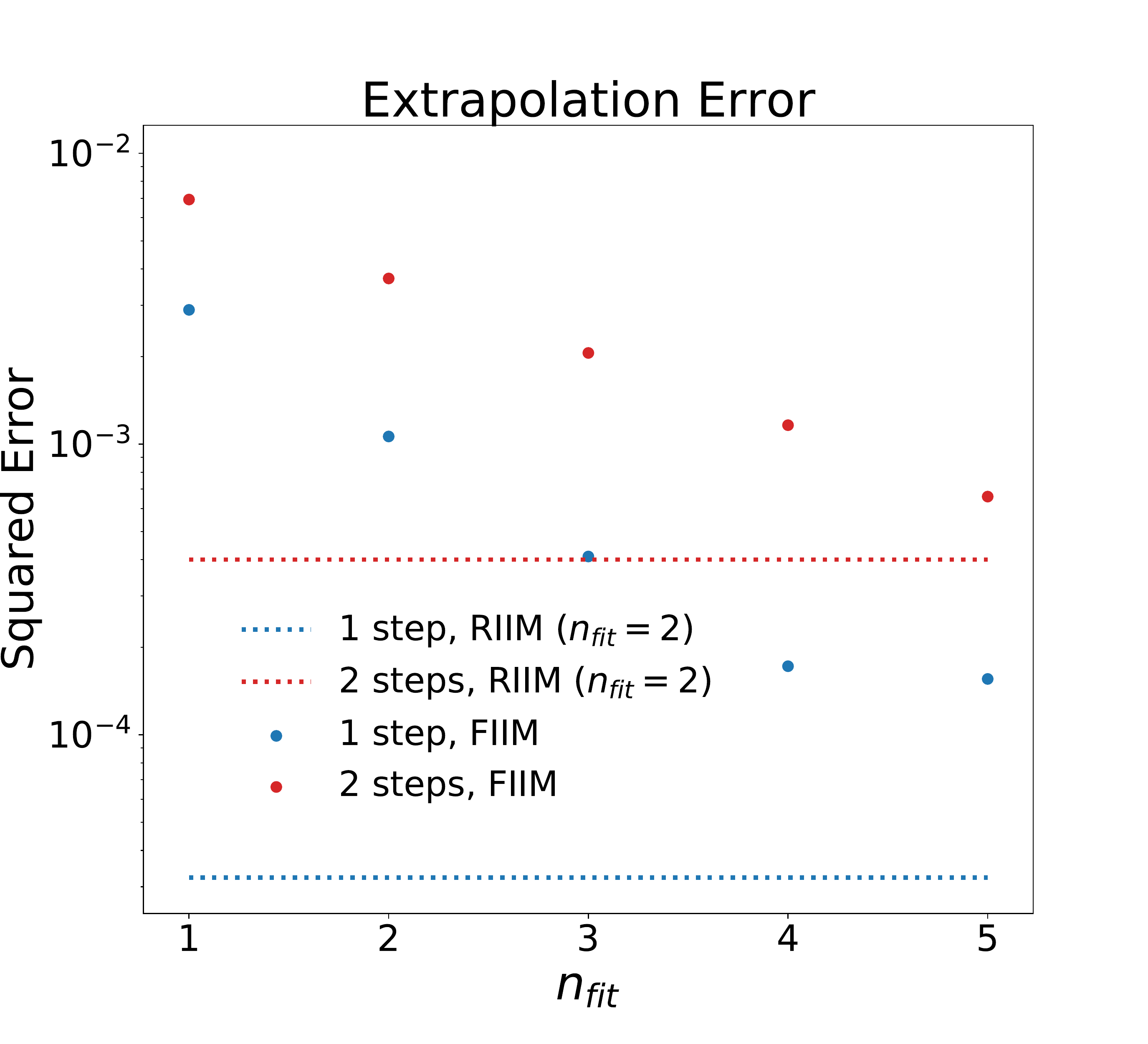}
\caption{The squared error of the observable value extrapolated using FIIM plotted against $n_\text{max}$. The dashed lines indicate the error using second-order RIIM. }
\label{fig:trotter_2}
\end{figure}
Fig.~\ref{fig:trotter_2} compares the error obtained from the FIIM and RIIM extrapolations over different values of $n_\text{max}$. The extrapolated error from RIIM up to $\mathcal{O}(\epsilon^2)$ is lower than any of the errors obtained through FIIM for all values of $n_\text{fit}$ in the 1-step case and in the 2-step case.  
%

\begin{comment}
\begin{itemize}
\item Compare the results for 3 different types of runs
\begin{itemize}
	\item Simulation with depolarizing CNOT noise
	\item Simulation with full Krauss CNOT noise
	\item Simulation with realistic noise model
	\item Runs on actual quantum hardware
\end{itemize}
\item For these, we should show results from the simple 2 CNOT circuit, as well as the harmonic oscillator
\end{itemize}
\end{comment}

%%%%%%%%%%%%%%%%%%%%%%%%%%%%%%%%%%%%%%%%%%%%%%%%%%%%%%%
\section{Conclusions}
\label{sec:conclusions}

We have performed a detailed study of zero noise extrapolation for correcting gate errors in quantum circuits.  The first aspect of this study was the formalization of the fixed identity insertion method (FIIM), which increases the circuit error by inserting pairs of gates after each \cnot~in the circuit.  This method has been studied in the past, but we derived analytic results for removing higher-order depolarizing noise.  These analytic results were previously known in the context of Hamiltonian evolution and are connected with the identity insertion formalism.  We also make the observation that these extended fits are equivalent to higher-order polynomial extrapolations.  

A key challenge with FIIM is that it requires a significant inflation in the gate count to achieve high precision.  We propose a new method whereby identities are randomly instead of deterministically inserted.  A careful choice of insertion probabilities can result in the same formal accuracy as FIIM but with far fewer gates [$(2n-1)n_\cnot$ versus $n_\cnot + 2(n-1)$].  This method will provide access to moderately deep circuits where FIIM is not applicable for near-term devices.

Finally, we have discussed the impact of other important sources of noise.  In particular, ZNE does not remove generic non-depolarizing noise.  Furthermore, large shot noise can spoil the high-order depolarizing noise cancellation.  New techniques may be required to mitigate these sources of noise within the ZNE framework.

In the era of NISQ hardware, zero noise extrapolation will continue to play an important role for enhancing the precision of quantum algorithms.  Identity insertions provide a practical error-model agnostic and software-based approach for enhancing errors in a controlled way.  The new RIIM method has extended this methodology for finer control over the error scaling and will extend the efficacy of zero noise extrapolation to moderate-depth circuits.   Combined with readout error mitigation, these techniques will provide a complete package for improving the accuracy of near term calculations on quantum devices. 

\begin{table*}[h!]
\begin{tabular}{c|ccccccccccccc}
$n_{\rm max}$ &  $a_{\{3\}}$ & $a_{\{5\}}$ & $a_{\{3,3\}}$ & $a_{\{7\}}$ & $a_{\{5,3\}}$ & $a_{\{3,3,3\}}$ & $a_{\{9\}}$ & $a_{\{7 ,3\}}$  & $a_{\{5 ,5\}}$ & $a_{\{5,3,3\}}$ & $a_{\{3, 3,3, 3\}}$
\\\hline
1 & $-\frac{1}{2}$ &&&&&&&&
\\\hline
2 & $-\frac{N_{\rm c} + 4}{4}$ & $\frac{3}{8}$ & $\frac{1}{4}$ &&&&&&
\\\hline
3 & $-\frac{N_{\rm c}^2 + 10 N_{\rm c} + 24}{16}$ & $\frac{3(N_{\rm c} + 6)}{16}$ & $\frac{N_{\rm c} + 6}{8}$ & $-\frac{5}{16}$ & $-\frac{3}{16}$ & $-\frac{1}{8}$ &&&
\\\hline
4 & $-\frac{N_{\rm c}^3 + 18 N_{\rm c}^2 + 104 N_{\rm c} + 192}{96}$ & $\frac{3N_{\rm c}^2 + 32 N_{\rm c} + 154}{64}$ & $\frac{N_{\rm c}^2 + 14 N_{\rm c} + 59}{32}$ & $-\frac{45}{32}$ & $-\frac{3 N_{\rm c} + 29}{32}$ & $- \frac{N_{\rm c}+8}{16}$ &  $\frac{35}{128}$  & $0$ & $\frac{29}{64}$ & $\frac{3}{32}$ & $\frac{1}{16}$
\end{tabular}
\caption{Table giving the coefficients for higher order RIIM fits.}
\label{RIIMCoefficients}
\end{table*}

\begin{acknowledgments}

We would like to thank Andrew Christensen, Yousef Hindy, Mekena Metcalf, John Preskill, Miro Urbanek, and Will Zeng for useful discussions.  This work is supported by the U.S. Department of Energy, Office of Science under contract DE-AC02-05CH11231. In particular, support comes from Quantum Information Science Enabled Discovery (QuantISED) for High Energy Physics (KA2401032) and the Office of Advanced Scientific Computing Research (ASCR) through the Accelerated Research for Quantum Computing Program.  

\end{acknowledgments}

\bibliographystyle{apsrev4-1}
\bibliography{myrefs}

\end{document}